\documentclass{webofc}
\usepackage[varg]{txfonts}   
\newcommand{\be}{\begin{equation}}
\newcommand{\BE}[1]{\begin{equation} \n{#1}}
\newcommand{\ee}{\end{equation}}
\newcommand{\ba}{\begin{eqnarray}}
\newcommand{\ea}{\end{eqnarray}}
\newcommand{\eq}[1]{(\ref{#1})}
\newcommand{\n}[1]{\label{#1}}

\newcommand{\hh}{,\hspace{0.5cm}}
\newcommand{\hhh}{,\hspace{0.2cm}}
\newcommand{\pa}{\partial}
\newcommand{\ins}[1]{{\mbox{\tiny #1}}}

\newcommand{\hor}{\stackrel{ {\mbox{\tiny H}}}{=} }

\begin{document}
\title{Remarks on non-singular black holes}
%
%

\author{\firstname{Valeri P.} \lastname{Frolov}\inst{1}\fnsep\thanks{\email{vfrolov@ualberta.ca}}
}

\institute{Theoretical Physics Institute, Department of Physics\\
University of Alberta, Edmonton, Alberta, Canada T6G 2E1 }

\abstract{We briefly discuss non-singular black hole models, with the main focus on the properties of non-singular evaporating black holes. Such black holes possess an apparent horizon, however the event horizon may be absent. In such a case, the information from the black hole interior may reach the external observer after the complete evaporation of the black hole. This model might be used for the resolution of the information loss puzzle. However, as we demonstrate, in a general case the quantum radiation emitted from the black hole interior, calculated in the given black hole background, is very large. This outburst of the radiation is exponentially large for models with the redshift function $\alpha=1$. We show that it can be suppressed by including a non-trivial redshift function. However, even this suppression is not enough to guarantee self-consistency of the model. This problem is a manifestation of a general problem, known as the "mass inflation". We briefly comment on possible ways to overcome this problem in the models of non-singular evaporating black holes.
}
\maketitle
\section{Introduction}
\label{intro}

General Relativity predicts existence of black holes. Hundred years after the creation of this theory this prediction was confirmed by LIGO discovery of the gravitational waves emitted in the black hole coalescence. The general relativity is the most reliable theory of the gravity. However, it has its own problems, connected with inevitable existence of singularities. According to this theory such singularities should exist inside black holes. In the well known exact solutions describing stationary vacuum isolated black holes these singularities are connected with the infinite growth of the curvature invariants, such as the square of the Weyl tensor. This means that anisotropy of the spacetime infinitely grows in the vicinity of such singular points. This behavior is similar to the case of the contracting Kasner universe. In order to escape the black hole singularity one needs to find a mechanism suppressing the infinite growth of the anisotropy.This problem is more severe than the  singularity problem in homogeneous isotropic cosmological models, where it is sufficient to suppress the growth of the Ricci tensor.

In order to cure this “disease” several modifications of the gravity theory were proposed. For example, effects of vacuum polarization and particle creation in the black hole interior result in the additional terms in the effective action for gravity. These modifications would contain both, higher derivatives and non-locality. One can also modify the fundamental gravity equations, by considering, for example, a theory with higher derivatives, $f(R)$ and Gauss-Bonnet theories, etc. More recently, it was proposed a special non-local modification of the Einstein theory, known as a ghost free gravity. At the more fundamental level, one can treat gravity as an emergent phenomenon (string, loops, etc.). However, in spite of many efforts, till now there is no generally accepted mechanism of black hole singularities removal. In the absence of the adopted theory it is instructive to discuss what kind of modifications one could expect when gravity is UV complete. Such an analysis can be fruitful only if some "natural" assumptions concerning the properties of such "full" theory are imposed.

In this talk I briefly discuss such a 'phenomenological' approach, describe proposed models of a non-singular black hole and their properties.
The main new feature of such models is that instead of the event horizon they might have a closed apparent horizon. As a result, the information sent inside the black hole can return to external space after the black hole complete evaporation. This opens an interesting possibility for solving the so-called information paradox. I shall also discuss the quantum effects in the interior of a non-singular black hole and demonstrate that huge out-burst of the quantum radiation from the interior should be expected, unless one specially adapts the metric near the inner horizon.  The main original results, presented in this talk, are based on the following publications \cite{Frolov:2016pav,Frolov:2016gwl,Frolov:2017rjz}.

\section{Phenomenological approach and the limiting curvature conjecture}
\label{sec-1}

One of the main assumptions in the phenomenological description of non-singular black holes is that there exists a critical energy $\mu$ and the corresponding length-scale parameter $\ell=\mu^{-1}$. The metric should be modified when the spacetime curvature ${\cal R}$ becomes comparable with $\ell^{-2}$. At the same time we assume that one can use the classical metric $g_{\mu\nu}$, which is a solution of the (unknown) modified gravity equations. In other words, the length scale, where quantum gravity effects become important, is much smaller than $\ell$. We are looking for black-hole metrics which do not have curvature singularities. The first model of a non-singular black hole was proposed by Bardeen \cite{Bardeen:1968_a}, who considered a collapse of a charged matter with a charged matter core inside the black hole instead of its singularity. A model of a non-singular evaporating black hole was first discussed in \cite{Frolov:1979tu,Frolov:1981mz}.

Different models of neutral, charged and rotating non-singular black holes were proposed and discussed later \cite{Dymnikova:1992ux,Borde:1996df,AyonBeato:1998ub,Lemos:2011dq,Uchikata:2012zs,Flachi:2012nv,DeLorenzo:2014pta,
Balart:2014cga,Ghosh:2014hea,DeLorenzo:2015taa,Haggard:2014rza,Lemos:2016ulj}. A general review of different models of non-singular black holes and additional references on this subject can be found in \cite{2008arXiv0802.0330A,2014arXiv1406.4098B,Frolov:2016pav,2017arXiv170609204B,Chinaglia:2017uqd}.

In a general case, a regular solution besides some critical scale parameter $\ell$, which is a parameter of the corresponding UV complete theory, contains also such parameters as mass and charge, which specify a concrete solution. The regularity of the solution means that for a fixed value of these parameters the curvature of the spacetime is finite. However, in a general case there is no guarantee that the maximal value of the curvature would not infinitely grow when, say, the mass (and/or charge) takes special value. It is reasonable to assume that for a viable fundamental theory the absolute maximal value of the curvature is restricted by some fundamental value. In other words, the curvature invariants are uniformly restricted by some universal value
$|{\cal R}|\le c\ell^{-2}$, where $c$ is a dimensionless constant which depends only on the type of the curvature invariant.
This assumption, called the limiting curvature conjecture, was proposed in  \cite{Markov:1982,Markov:1984ii,Polchinski:1989ae}.

\section{Non-singular black hole models}
\label{sec-2}

A general   static spherically symmetric metric  in a four-dimensional spacetime  can be written in the form
\be\n{a.1}
ds^2=-f \alpha^2 dv^2+2 \alpha dv dr+r^2 d\omega^2\, ,
\ee
where  $f=f(r)$ and $\alpha=\alpha(r)$ are two arbitrary functions. This metric has the Killing vector $\xi^{\mu}\pa_{\mu}=\pa_v$. It is easy to see that
\BE{a.1b}
f=(\nabla r)^2\hh f \alpha^2 =-\xi^2\, .
\ee
We call $\alpha$ a redshift function.
In a spacetime with a horizon, $f(r)$ vanishes at the position $r_H$ of the apparent horizon. For a regular static metric such a horizon is at the same time the Killing horizon, so that the redshift function $\alpha(r_H)$ is finite there.

If the metric has a horizon where $f(r_H)=0$ then
\be\n{a.1a}
\xi^{\nu}\xi^{\mu}_{\ ;\nu}\hor \kappa \xi^{\mu}\hh \kappa={1\over 2}(\alpha f')|_{r=r_H}\, .
\ee
By definition, $\kappa$ is the surface gravity. The value of $\kappa$ depends on the choice of the normalization of the Killing vector.
In an asymptotically flat spacetime one usually puts $\xi^2|_{r=\infty}=-1$.  A conditions that there is no solid angle deficit implies $f|_{\infty}=1$. Hence one also has $\alpha|_{\infty}=1$.

Let $R$ be the Ricci scalar, $S_{\mu\nu}=R_{\mu\nu}-{1\over 4}g_{\mu\nu}R $, and $C_{\mu\nu\alpha\beta} $ be the Weyl tensor. Let us define the following quadratic in the curvature invariants ${\cal S}^2=S_{\mu\nu} S^{\mu\nu}$ and ${\cal C}^2=C_{\mu\nu\alpha\beta} C^{\mu\nu\alpha\beta}$. Then one has
\ba\n{a.5}
R&=& f''+{4\over r}f'-2{f-1\over r^2}+{1\over \alpha}\left(2f\alpha''+3 f' \alpha' +{4\over r} f \alpha'\right) \, ,\\
{\cal C}&=&{1\over \sqrt{3}}\left[ f'' -{2\over r}f'+2{f-1\over r^2}+ {1\over \alpha}\left(2f \alpha''+3 f' \alpha' -{2\over r} f \alpha'\right) \right]\, .
\ea
An expression for ${\cal S}$ is similar but quite long and we do not present it here.

We assume that the metric \eq{a.1} is finite at the origin $r=0$, so that
\be
f=f_0+f_1 r+f_2 r^2+O(r^3)\hh
\alpha=\alpha_0+\alpha_1 r+\alpha_2 r^2+O(r^3)\, .\n{a.3}
\ee
In a general case the curvature invariants $R$ and ${\cal C}$ are singular at the origin and have divergences $\sim r^{-2}$ and $\sim r^{-1}$. Conditions that these divergences are absent and the geometry is regular are
$f_0=1$ and $f_1=\alpha_1=0$. Under these assumptions one has
\be\n{a.3b}
R=-12\left( f_2+{\alpha_2\over \alpha_0}\right)+O(r^2)\hh
{\cal S}=2\sqrt{3} {\alpha_2\over \alpha_0}+O(r^2)\hhh
{\cal C}=O(r^2)\, .
\ee

\section{Hayward model}

Let us consider first the case when $\alpha=1$. We assume  also that $f$ is a rational function of $r$
\BE{a.7b}
f(r)={P_n(r)\over \tilde{P}_n(r)}\, ,
\ee
where $P_n$ and $\tilde{P}_n$ are polynomials of the order $n>1$. One an check  that for $n\le 2$ these  metrics cannot be both asymptotically flat at infinity and regular at the center (see \cite{Frolov:2016pav}). However, for $n=3$ such metrics exist. An example is the following  a non-singular black hole metric was proposed by \cite{Hayward:2005gi} and discussed in \cite{Frolov:2014jva}
\be\n{a.8}
f=1-{2M r^2\over r^3 +2M \ell^2}\, .
\ee
This function has the following asymptotics
\BE{a.8a}
f=1-{2M\over r}+O(r^{-4})\hh
f=1-{r^2\over \ell^2}+O(r^5).
\ee

\begin{figure}[h]
\centering
\includegraphics[height=4.0cm]{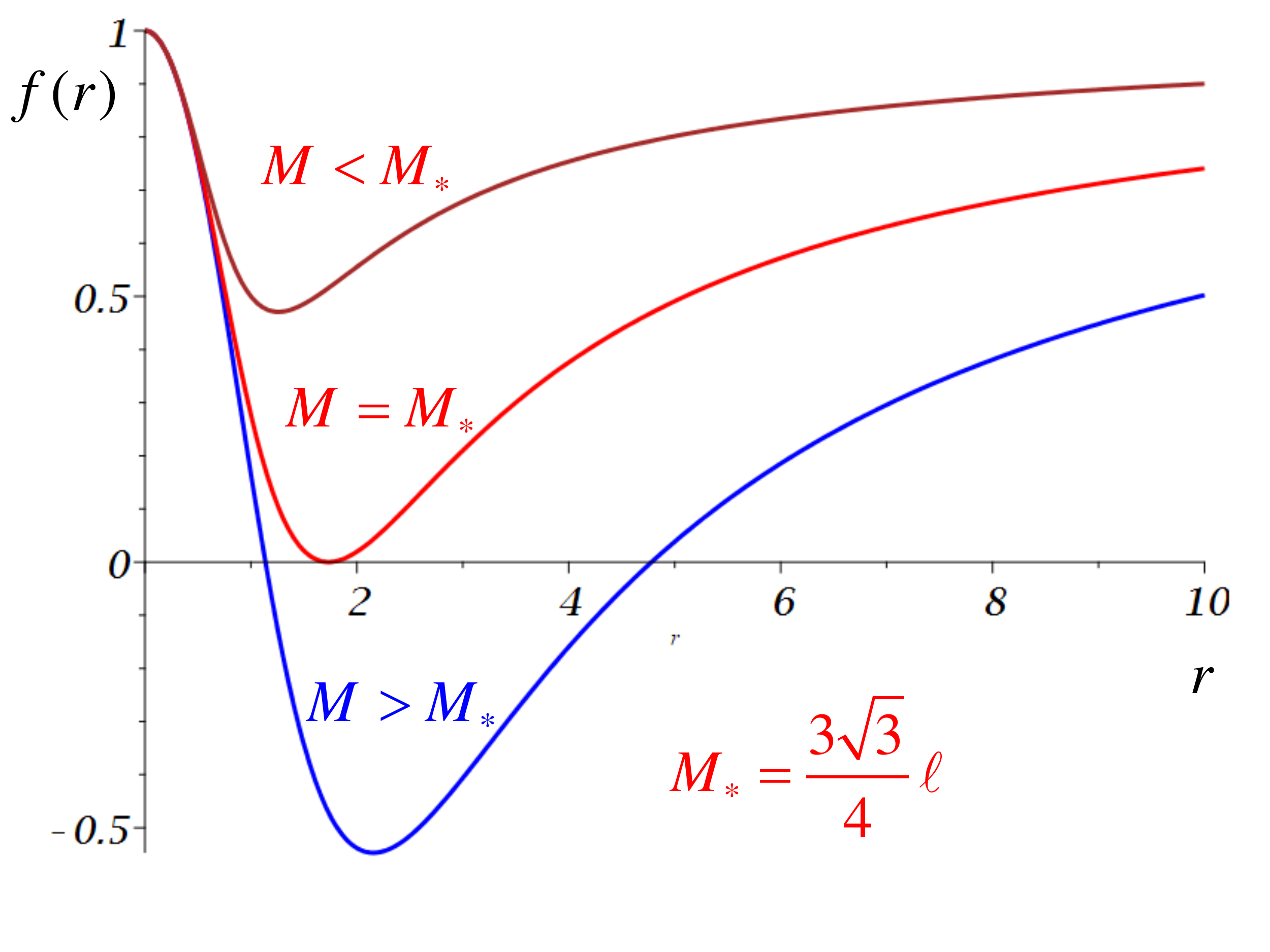}
\caption{Metric function $f(r)$ for different values  of the mass $M$. Its takes value 1 at $r=0$ and $r=\infty$. For $M<M_*={3\sqrt{3}\over 4}\ell$ this function is always positive and the metric does not have an apparent horizon. For $M>M_*$ it has two roots, corresponding to outer and inner apparent horizon.}
\label{F1}       
\end{figure}

Let us denote $r=(2M\ell^2)^{1/3}z$ and $B=(2M/\ell)^{2/3}$. Then one has
\be\n{a.10}
f=1-B{z^2\over z^3+1}\hhh
R=-{6\over  \ell^2} {z^3-2\over (z^3+1)^3}\hhh
{\cal S}={9\over  \ell^2} {z^3\over (z^3+1)^3}\hhh
{\cal C}={\sqrt{12}\over \ell^2} {z^3(z^3-2)\over (z^3+1)^3}\, .
\ee
All the curvature invariants, besides a common scale factor $\ell^{-2}$, contain rational functions of the variable $z$. The denominators of these functions are positive, while the functions themselves vanish at infinity and are either finite or vanish at $z=0$. Thus, their absolute value is restricted by some constant, independent of the parameter $M$. Hence, the metric (\ref{a.8}) obeys the limiting curvature condition.

\section{Non-singular evaporating black holes}

\subsection{Sandwich model}

In order to describe an evaporating non-singular black hole one can put the mass parameter $M$ in the metric (\ref{a.8}) to be time-dependent, $M=M(v)$. This function should be chosen so that it correctly reproduces the black hole mass decrease in the process of the Hawking evaporation. We consider this case later. It is instructive to discuss first a simplified version of the metric, describing a so-called sandwich black hole. Namely, we assume that a black hole is formed as a result of a spherical collapse of a null shell of mass $M$, it exists for time $\Delta v$, and after disappears, as a result of the collapse of another null shell of mass $-M$. During the time interval $\Delta v$ the metric coincides with  (\ref{a.8}) \cite{Frolov:2016gwl}. Certainly, this model is quite different from an expected behavior of the evaporating black hole. However, they have the following common property: a finite time of the `black hole'' existence.

The metric function \eq{a.8} between the shells contains 2 parameters, the mass $M$ and the "fundamental length" $\ell$. In what follows we assume that the mass $M$ is large enough, so that the function $f$ has two positive zeros $r_{1,2}$. In such a case it is possible to use $r_{1,2}$ as two new parameters, instead of $M$ and $\ell$. We also use the following notation $p=r_1/r_2$. Then the metric function $f(r)$ takes the form
\be\n{a.5a}
f(r)={(r-r_{1})(r-r_{2})(r-r_0)\over r^3 -r_{1} r_{2} r_0}\hh \quad
r_{1}=p\hh r_{2}=1\hh r_0=-{p\over p+1}\, .
\ee
The parameter $p$ is a position of the outer horizon in the dimensionless units, while  the inner horizon is located at $r=1$. The quantity $r_0$ is negative. The metric \eq{a.5a} is uniquely specified by two quantities, the mass parameter $p$ and the duration of the black hole existence, $q=\Delta v/r_2$. Because of its rather simple form, many results concerning the properties of such a spacetime can be obtained in an analytical form.
The dimensionless surface gravities, calculated for each of these horizons, are
\be\n{a.7}
\kappa_{1}={(p-1)(p+2)\over 2 p(p^2+p+1)}\hhh
\kappa_{2}=-{(p-1)(2p+1)\over 2(p^2+p+1) }\hhh
\kappa_0={ (p+1)(p+2)(2p+1)\over 2 p(p^2+p+1)} .
\ee

\begin{figure}[h]
\hfill
\includegraphics[height=6.0cm]{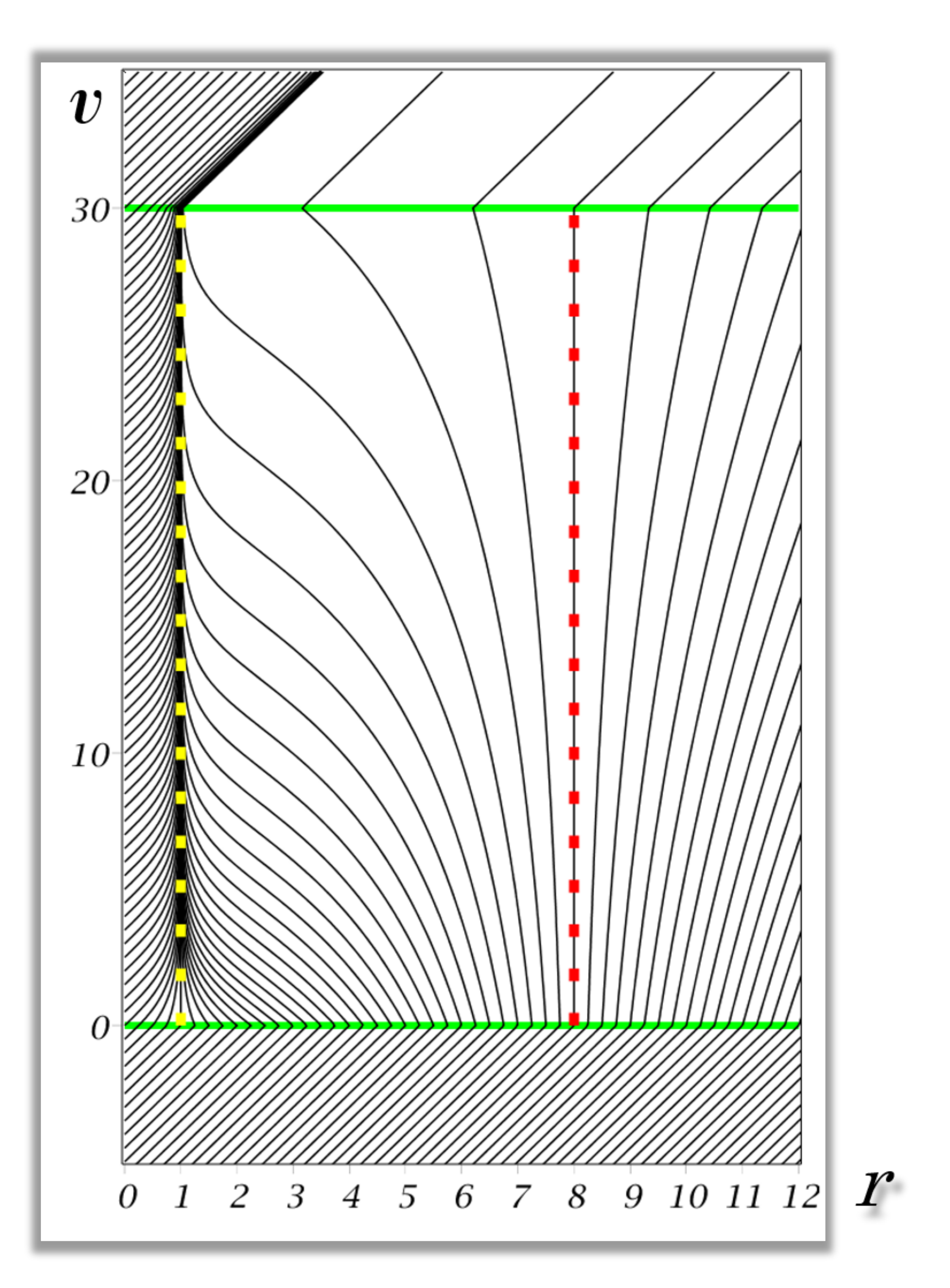}
\hspace{1cm}
\includegraphics[height=6.0cm]{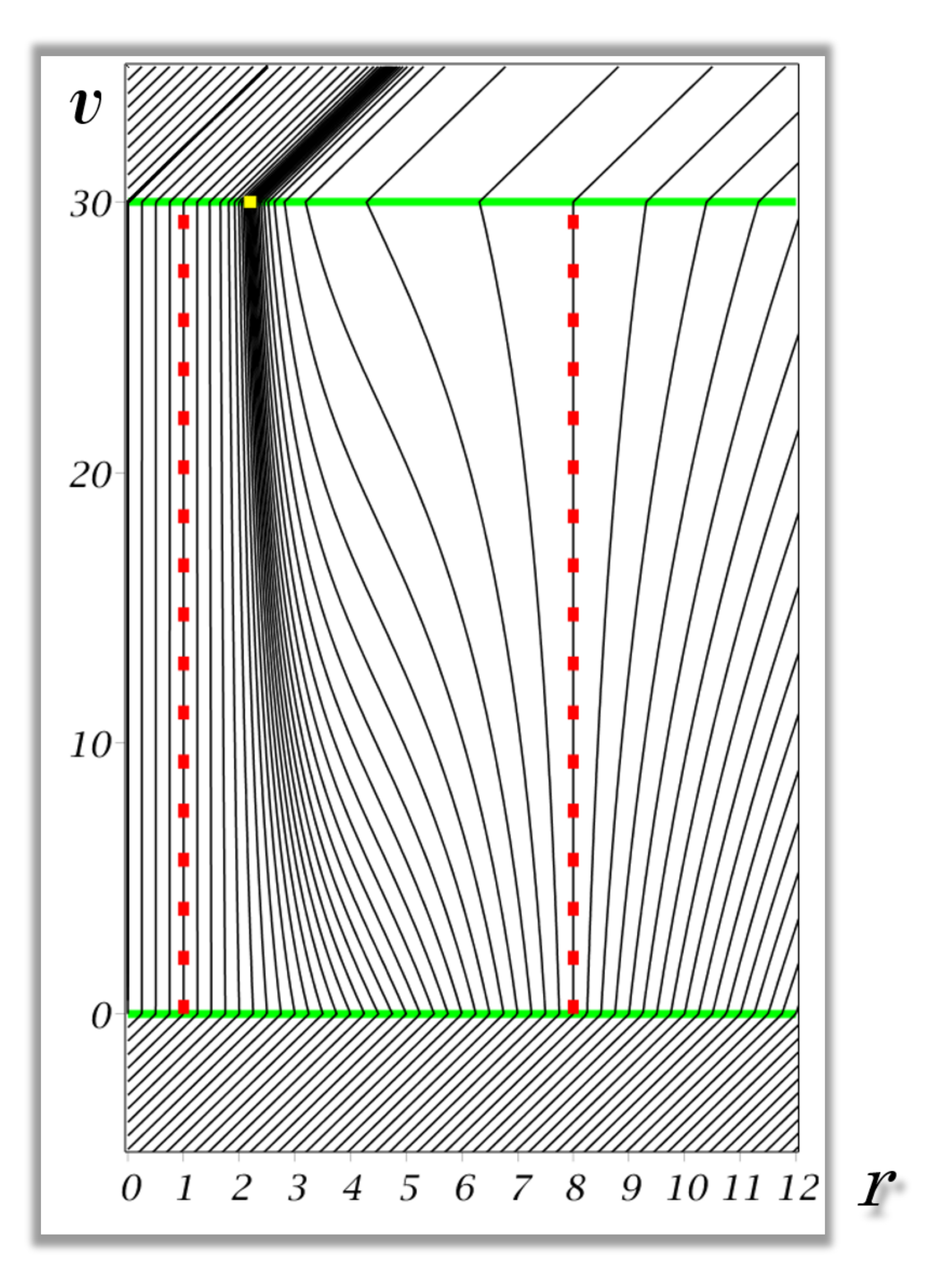}
\hspace{2cm}
\caption{Null rays propagation in the spacetime of a sandwich black hole. The plots are constructed for $p=8$ and $q=30$. The left plot illustrates the null rays in the metric with $\alpha=1$, while the right plot is for $\alpha={r^6+1\over r^6+1+p^4}$.}
\label{F2}       
\end{figure}

The motion of the incoming radial null rays in this geometry is rather simple. They are described by the equation $v=$const. Such rays pass through the center $r=0$ and become outgoing. Out-going null rays in the sandwich black-hole geometry are shown at Figure~\ref{F2}. The left plot in the Figure~\ref{F2} shows the null rays in the sandwich geometry with $\alpha=1$.
The rays that cross the first null shell with $r<1$ propagate near the origin, being accumulated near the inner horizon from its inner side. The rays  that cross the first null shell between $r=1$ and $r=p$ are also attracted to the inner horizon and are accumulated near it from its outer side. Those rays  that cross the first shell at $r>p$ are propagating away from the outer horizon. All incoming rays with the origin at $0<v<q$, after passing the inner horizon cross the center $r=0$ and returns to inner horizon from inside. They   are also accumulated in the narrow domain in its vicinity. In other words, the inner horizon plays the role of an attractor for the outgoing rays, while the outer horizon repulses the rays. After the collapse of the second shell with a negative mass, all the outgoing null rays freely propagate to the future null infinity ${\cal I}^+$. For $q\gg p$ the beam of out-going null rays propagating near the inner horizon is highly squeezed. The corresponding squeezing factor at time $v=q$ is $\sim \exp(-q \kappa_2)\sim \exp(q)$.

The right plot at the Figure~\ref{F2} illustrates the propagation of null rays in the sandwich black hole geometry with a non-trivial redshift function $\alpha$. We choose it in the form $\alpha=(r^6+1)/(r^6+1+p^4)$.
For large $p$ this factor near the center and the inner horizon $\alpha_0\sim p^{-k}$ is small. Since $\kappa_2\sim -\alpha_0$, the surface gravity of the inner horizon is greatly reduced. As a result, it takes longer time to accumulate the null rays near the inner horizon.

\subsection{Smooth mass function}

At first, we assume that $\alpha=1$, and put
\be\n{a.3}
f=1-{2M(v) r^2\over r^3 +2M(v) \ell^2+\ell^3} .
\ee
In the static case this metric function differs from \eq{a.8} by the presence of an extra $\ell^3$ term in the denominator. Such a modification makes the geometry near the point $r=0$ in the limit $M\to 0$ more smooth.
Since now the position of the inner horizon depends on time, it is more convenient to use $\ell$ as a scale parameter. For this choice $\ell=1$
\be\n{aa}
f=1-{\mu(v) r^2\over r^3 +\mu(v)\,+1} \hh \mu(v)\equiv 2M(v).
\ee

For sufficiently large $\mu$ this metric describes the nonsingular evolving black hole with two apparent horizons: the outer horizon $r_1(v)$ and the inner one $r_2(v)$. Their radii are determined by the condition $f=0$
and can be written explicitly ($\mu=\mu(v)$)
\be
r_\ins{1}={1\over 2}\left(\mu+1+\sqrt{\mu^2-2\mu-3}\right)\hh
r_\ins{2}={1\over 2}\left(\mu+1-\sqrt{\mu^2-2\mu-3}\right)\, .
\ee
The third root of the cubic equation $f=0$ corresponds to a negative value of radius $r_\ins{0}=-1$ and is not relevant for our study.
The inner and outer apparent horizons exist only when $\mu(v)> 3$.
One can easily calculate the surface gravity on the outer and inner horizons
\be
\kappa_\ins{1,2}=\mu{r_\ins{1,2}(r_\ins{1,2}^3-2\mu-2)\over2(r_\ins{1,2}^3+\mu+1)^2}.
\ee
One can check that, provided horizons exist,  $\kappa_\ins{1}\ge 0$, while $\kappa_\ins{2}\le 0$.
For large masses $\mu\gg 3$ we have
\be
r_\ins{1}=\mu-{1\over\mu}+O(\mu^{-2})\, ,
r_\ins{2}=1+{1\over\mu}+O(\mu^{-2})\, ,
\kappa_\ins{1}={1\over 2\mu}-{1\over \mu^3}+O(\mu^{-4})\, ,
\kappa_\ins{2}=-1+{5\over 2\mu}+{1\over \mu^3}+O(\mu^{-4}).
\ee
For $\mu= 3$ the horizons merge and we obtain an extremal case with $r_\ins{1}=r_\ins{2}=2$ and $\kappa_\ins{1}=\kappa_\ins{2}=0$. For $\mu<3$ the metric has no the apparent horizon.

\begin{figure}[h]
\centering
\includegraphics[height=4.0cm]{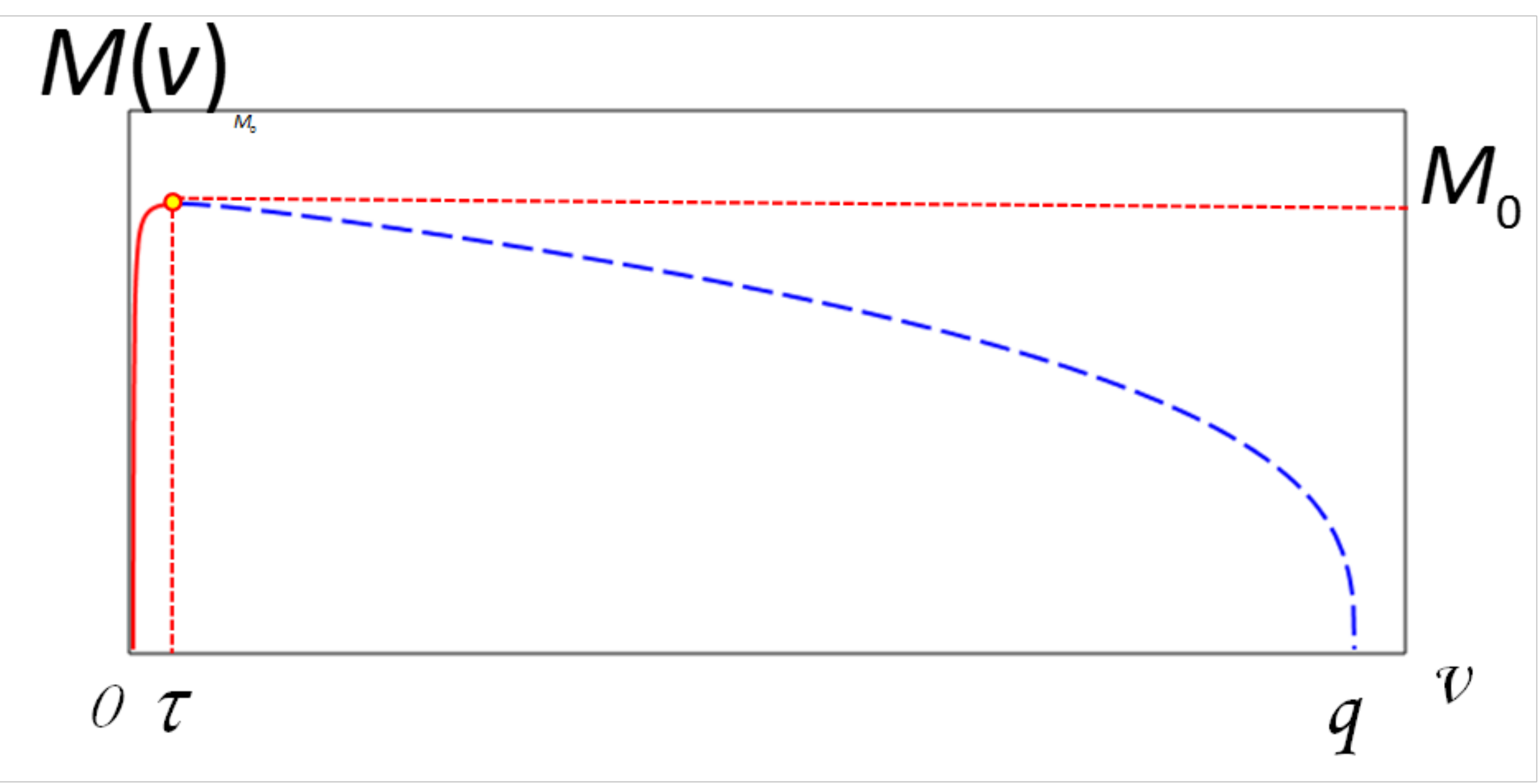}
\caption{This plot schematically illustrates a typical shape of the mass function $\mu(v)=2M(v)$. The dashed line describes the stage of the Hawking evaporation of the black hole. }
\label{F3}       
\end{figure}

For concrete calculations we chose the function $\mu(v)$ in the form
\be\n{muv}
\mu(v)=\begin{cases}
\displaystyle{{2v(q-\tau)(q-v)^{1/3}\over C\big(3(q-\tau)v-\tau v+\tau^2\big)}}& 0\le v\le q,\\
0 &  v<0,~~\mbox{and}~~ v>q.
\end{cases}
\ee
It is schematically shown at Figure~\ref{F3}. This function has a maximum at $v=\tau$, where $\mu(\tau)=\mu_0$. During the interval of $v$ from 0 to $\tau$ the mass parameter $\mu$ grows from 0 to $\mu_0$. This is a phase of the null fluid collapse which results in the formation of a black hole. Starting from $v=\tau$ the mass reduces as a result of the Hawking radiation. At this stage $\dot{\mu}\sim \mu^{-2}$. This phase continues up to the time $v=q$
\be
q=(C\mu_0)^3+\tau\, .
\ee
The parameter $C$ depends on the number of distinct polarizations of particles which are emitted.
Note that overall picture of quantum particle creation depends mostly on the asymptotic properties of the function $\mu(v)$ rather than on variations of its shape.

\begin{figure}[h]
\hfill
\includegraphics[height=6.0cm]{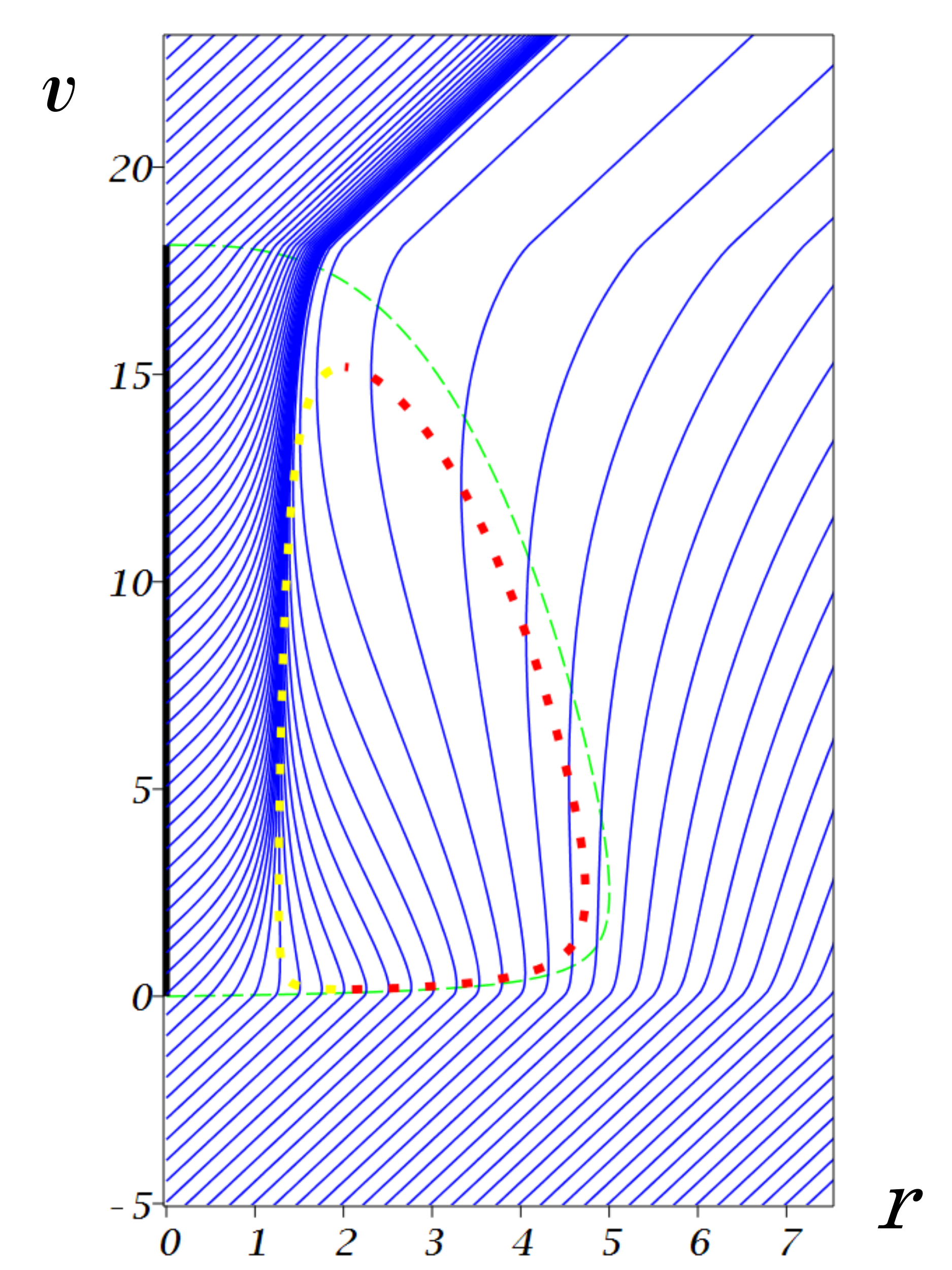}
\hspace{1cm}
\includegraphics[height=6.0cm]{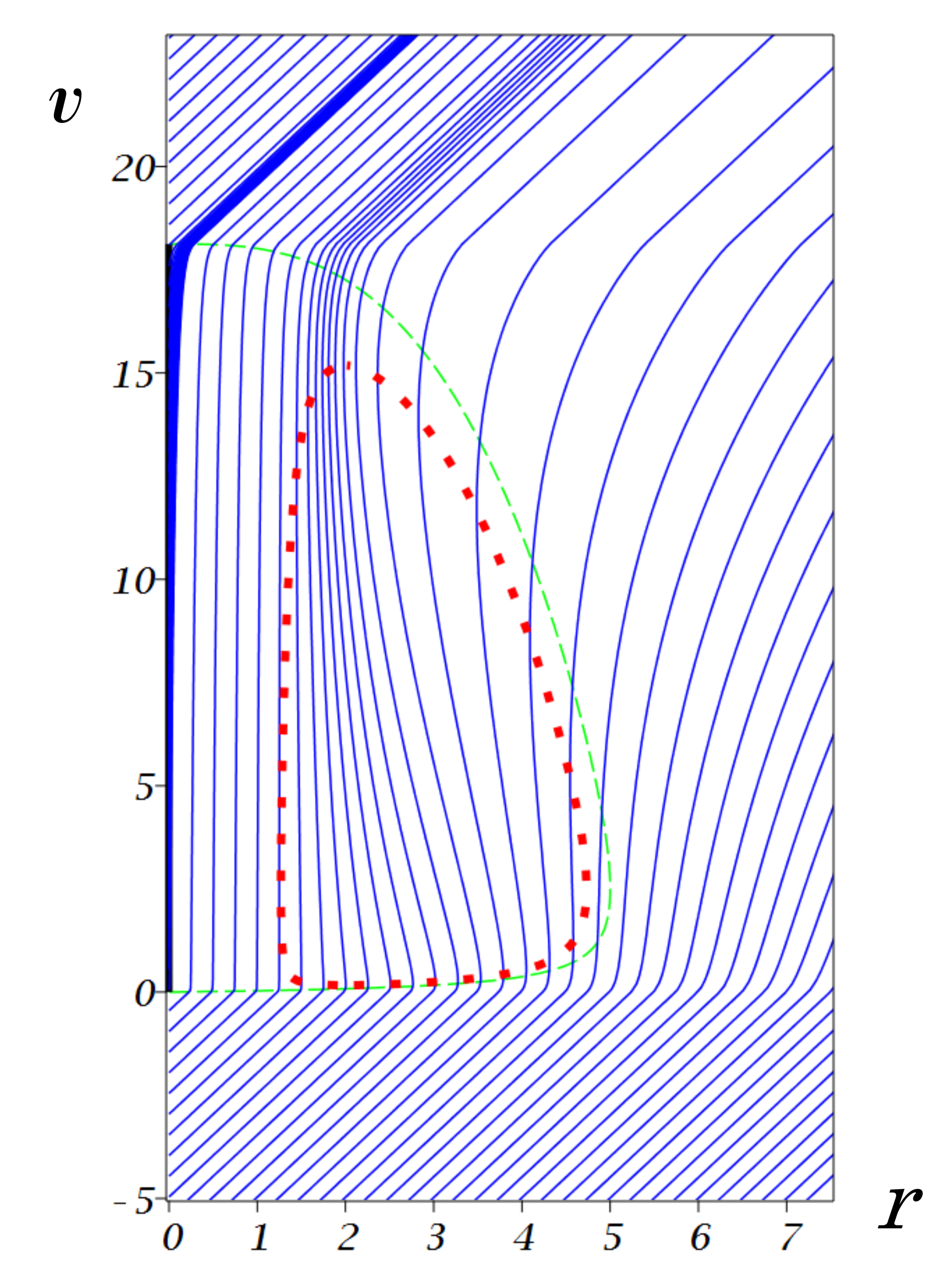}
\hspace{2cm}
  \caption{Null rays propagation in the spacetime of a non-singular evaporating black hole. The plots are constructed for the value of the parameters $\mu_0=5$, $\tau=2.5$ and $C=1/2$. The left plot is for the case $\alpha=1$, while the right plot is for $\alpha=(1+r^5)/(1+r^5+\mu^3)$.
\label{F4}}
\end{figure}

The left plot of Figure~\ref{F4} shows the null rays in the geometry of an evaporating non-singular black hole with $\alpha=1$. The motion of the incoming radial null rays in this geometry is quite simple. They are described by the equation $v=$const. Such rays pass through the center $r=0$ and then become outgoing. The incoming null rays with $v<0$ which cross the collapsing null fluid shell at small radius, $0<r_-<1$, are focussed to the inner horizon from its interior. Similar rays for small $r>1$ are also focussed, but in the exterior of the inner horizon. As a result the corresponding beams of light are squeezed by the factor $\sim \exp(-\kappa_2 q)$, where $q$ is the life-time of the black holes. For the specially chosen $\alpha$ this squeezing is relaxed. The corresponding plot is shown in the right picture at Figure~\ref{F4}.  Let us emphasize, that qualitatively the character of the null ray behavior in this geometry is quite similar to the sandwich black hole case.

\section{Quantum effects}

\subsection{Radial null rays and  ${\cal I}^-\to{\cal I}^+$ map}

In the chosen coordinates, radial incoming null rays are described by the relation $v=$const , while the outgoing null rays satisfy the following equation
\be\n{outr}
{dr\over dv}={1\over 2} \alpha f\, .
\ee
Consider an outgoing radial null ray which reaches the future null infinity ${\cal I}^+$ at the retarded time $u=u_+$ and trace it backward in time. We denote by $u_-$ time $v$ when it crosses the center $r=0$. After this it propagates to the past null infinity ${\cal I}^-$ at  $v=u_-$. The relation $u_-=u_-(u_+)$ establishes a map between ${\cal I}^+$ and ${\cal I}^-$. Since the tangent vector to the null ray trajectory is uniquely determined by equation (\ref{outr}), any two outgoing null rays cannot intersect. This means that $u_+(u_-)$ is a monotonically increasing function, so that $du_+/du_->0$. The map function contains the information concerning observables at ${\cal I}^+$. In particular, it allows one to calculate in 2D approximation the quantum energy flux at ${\cal I}^+$ from a non-singular black hole.

\subsection{The energy flux}

In the 2D approximation, the energy flux of massless particles, created from the initial vacuum state, is given by the following expression
\be\n{EEEE}
\dot{E}={dE\over du_+}={1\over 48\pi}\left[ -2{d^2 P\over du_+^2}+\left({d P\over du_+}\right)^2\right]\hh
P=\ln \left| {du_-\over du_+}\right|\, .
\ee
This relation directly follows from a general result obtained by Fulling and Christensen \cite{Christensen:1977jc} for the quantum average of the stress-energy tensor of a massless scalar field in two dimensions, reconstructed from the conformal anomaly. The same expression can be also obtained by the variation of the Polyakov effective action with respect to the metric \cite{Frolov:1984,Vilkovisky:1985}.

\begin{figure}[h]
\hfill
\includegraphics[width=4.0cm]{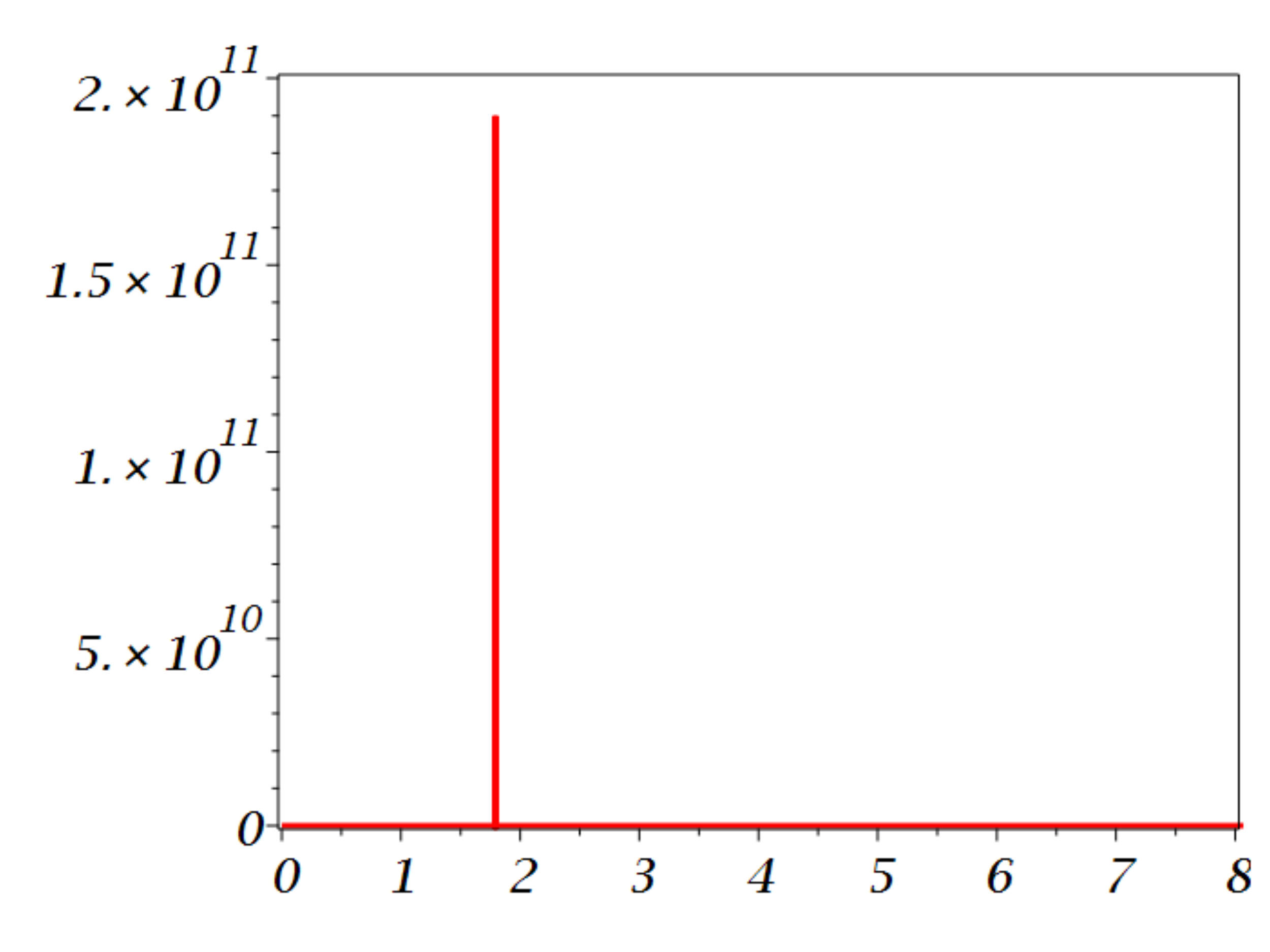}
\hspace{1cm}
\includegraphics[width=4.0cm]{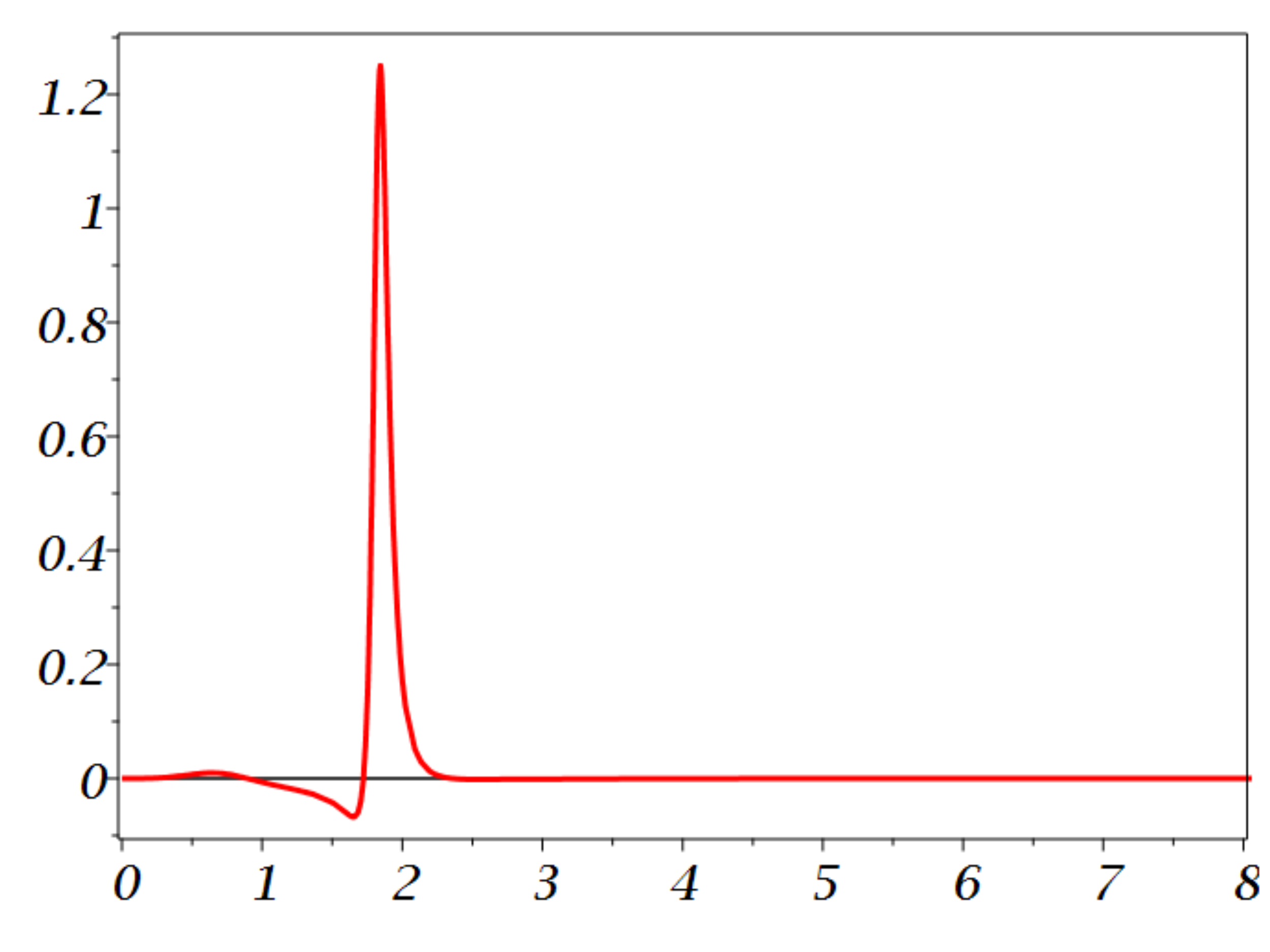}
\hspace{2cm}
\caption{Quantum energy flux $dE/du_+$ at the future null infinity from the inner horizon of an evaporating black hole. These plots are constructed for $\mu_0=3$, $\tau=6$. and $C=1/2$. $\alpha=1$ for the left plot and
$\alpha=(1+r^5)/(1+r^5+\mu^3)$ for the right plot, respectively. }
\label{F5}       
\end{figure}

For the sandwich-type black holes the map function can be found analytically (see \cite{Frolov:2016gwl}). For the smooth-type evaporating metric one needs to use numerical calculations. It is quite straightforward to calculate the map function. However, to obtain the energy flux one needs to find its  derivatives with respect time $u_+$ up to the third order. This makes this method not accurate. In order to overcome this problem, a new method, so called "bracket formalism", was developed in \cite{Frolov:2017rjz}. The main idea of this method is to consider a narrow beam of null rays near a fixed one and using the theory of perturbations to derive the equations for rays, propagating in the close vicinity to the chosen "fiducial" ray. It was shown that the obtained set of equations is greatly simplified if instead of usual derivatives to use their "bracket" generalizations, similar to the Schwarzian derivative.

The results of the calculations for sandwich and "realistic" models are quite similar. One can find all the details in \cite{Frolov:2016gwl,Frolov:2017rjz}. Here we just describe some of the main properties of the quantum energy flux. An external observer first registers rays emitted during the phase of creation of the black hole. For long enough life-time of the black hole, he/she will observe the Hawking radiation. This constant in time radiation is reproduced in our numerical calculations with very high accuracy. This stage is over when the rays from the black hole interior reach the observer. This part of the radiation is sensitive to details of the inner structure of the black hole, but the flux is relatively weak. Next what is seen by the observer is the outburst of radiation from the inner horizon. It intensity depends on the exponent of the factor
\be
-\mbox{[surface gravity of inner horizon]}\times \mbox{[black hole life time]}\, .
\ee
These fluxes for sandwich and "realistic" black hole models are similar. For illustration, we show this flux at Figure~\ref{F5} for "realistic models" with $\alpha=1$ and a specially chosen non-trivial $\alpha$.

\section{Discussion}

The outburst of the quantum energy flux from the inner horizon is related to a so-called mass inflation effect \cite{Poisson:1989zz,Poisson:1990eh} (see also \cite{Brown:2011tv}). The inner horizon has a negative surface gravity and it works as an attractor of the outgoing null rays. For a (quasi-) static case, the advanced time $v$ is the Killing time parameter and it differs from the affine parameter $\lambda$ along the outgoing null ray. Near the inner horizon they are related as follows $\lambda\sim \exp(-\kappa v)$. The momentum $p^{\mu}=dx^{\mu}/d\lambda$ of outgoing photons exponentially grows, while their energy, $E\sim -p_{v}$, remains constant because of the gravitational redshift. However, this compensation mechanism is broken when the position of the inner horizon moves. As a result, quanta leaving a black hole interior possess such exponentially large non-compensated blueshift.

For a special choice of the redshift function the exponential outburst of the radiation emitted  from the inner horizon can be suppressed. This happens when the surface gravity of the inner horizon is reduced. As a result, the time of the black hole evaporation is not sufficient for generation the exponential regime, and the energy flux decreases. The  energy flux for the model with $\alpha\ne 1$ it still  large. Partly this is connected with large contribution of the $\alpha$-anomaly. This contribution describes the particle creation in the incoming modes, which becomes large for fast change of the redshift function $\alpha$ during the formation of the black hole. As a result,  the model also violates a self-consistency requirement (see \cite{Bolashenko:1986mr}).
The obtained results indicate that the backreaction of the particles created  in the black hole interior should be properly taken into account. An interesting problem is a search for dynamical self-consistent non-singular black hole models.

\section*{Acknowledgments}

The authors thank the Natural Sciences and Engineering Research Council of Canada and the Killam Trust for their financial support.


\begin{thebibliography}{34}
\bibliographystyle{h-physrev4}

\bibitem{Frolov:2016pav}
V.P. Frolov, Phys. Rev. \textbf{D94}, 104056 (2016), \texttt{1609.01758}

\bibitem{Frolov:2016gwl}
V.P. Frolov, A.~Zelnikov (2016), \texttt{1612.05319}

\bibitem{Frolov:2017rjz}
V.P. Frolov, A.~Zelnikov, Phys. Rev. \textbf{D95}, 124028 (2017),
  \texttt{1704.03043}

\bibitem{Bardeen:1968_a}
J.M. Bardeen, \emph{{Non-singular general-relativistic gravitational
  collapse}}, in \emph{{Proceedings of of International Conference GR5,Tbilisi,
  USSR}} (1968), p. 174

\bibitem{Frolov:1979tu}
V.P. Frolov, G.A. Vilkovisky, \emph{{}}, in \emph{{The Second Marcel Grossmann
  Meeting on the Recent Developments of General Relativity (In Honor of Albert
  Einstein) Trieste, Italy, July 5-11, 1979}} (1979), p. 0455

\bibitem{Frolov:1981mz}
V.P. Frolov, G.A. Vilkovisky, Phys. Lett. \textbf{106B}, 307 (1981)

\bibitem{Dymnikova:1992ux}
I.~Dymnikova, Gen. Rel. Grav. \textbf{24}, 235 (1992)

\bibitem{Borde:1996df}
A.~Borde, Phys. Rev. \textbf{D55}, 7615 (1997), \texttt{gr-qc/9612057}

\bibitem{AyonBeato:1998ub}
E.~Ayon-Beato, A.~Garcia, Phys. Rev. Lett. \textbf{80}, 5056 (1998),
  \texttt{gr-qc/9911046}

\bibitem{Lemos:2011dq}
J.P.S. Lemos, V.T. Zanchin, Phys. Rev. \textbf{D83}, 124005 (2011),
  \texttt{1104.4790}

\bibitem{Uchikata:2012zs}
N.~Uchikata, S.~Yoshida, T.~Futamase, Phys. Rev. \textbf{D86}, 084025 (2012),
  \texttt{1209.3567}

\bibitem{Flachi:2012nv}
A.~Flachi, J.P.S. Lemos, Phys. Rev. \textbf{D87}, 024034 (2013),
  \texttt{1211.6212}

\bibitem{DeLorenzo:2014pta}
T.~De~Lorenzo, C.~Pacilio, C.~Rovelli, S.~Speziale, Gen. Rel. Grav.
  \textbf{47}, 41 (2015), \texttt{1412.6015}

\bibitem{Balart:2014cga}
L.~Balart, E.C. Vagenas, Phys. Rev. \textbf{D90}, 124045 (2014),
  \texttt{1408.0306}

\bibitem{Ghosh:2014hea}
S.G. Ghosh, S.D. Maharaj, Eur. Phys. J. \textbf{C75}, 7 (2015),
  \texttt{1410.4043}

\bibitem{DeLorenzo:2015taa}
T.~De~Lorenzo, A.~Giusti, S.~Speziale, Gen. Rel. Grav. \textbf{48}, 31 (2016),
  [Erratum: Gen. Rel. Grav.48,no.8,111(2016)], \texttt{1510.08828}

\bibitem{Haggard:2014rza}
H.M. Haggard, C.~Rovelli, Phys. Rev. \textbf{D92}, 104020 (2015),
  \texttt{1407.0989}

\bibitem{Lemos:2016ulj}
J.P.S. Lemos, V.T. Zanchin, Phys. Rev. \textbf{D93}, 124012 (2016),
  \texttt{1603.07359}

\bibitem{2008arXiv0802.0330A}
S.~{Ansoldi}, ArXiv e-prints  (2008), \texttt{0802.0330}

\bibitem{2014arXiv1406.4098B}
J.M. {Bardeen}, ArXiv e-prints  (2014), \texttt{1406.4098}

\bibitem{2017arXiv170609204B}
J.M. {Bardeen}, ArXiv e-prints  (2017), \texttt{1706.09204}

\bibitem{Chinaglia:2017uqd}
S.~Chinaglia, S.~Zerbini, Gen. Rel. Grav. \textbf{49}, 75 (2017),
  \texttt{1704.08516}

\bibitem{Markov:1982}
M.~Markov, JETP Letters \textbf{36}, 265 (1982)

\bibitem{Markov:1984ii}
M.~Markov, Annals Phys. \textbf{155}, 333 (1984)

\bibitem{Polchinski:1989ae}
J.~Polchinski, Nucl.Phys. \textbf{B325}, 619 (1989)

\bibitem{Hayward:2005gi}
S.A. Hayward, Phys.Rev.Lett. \textbf{96}, 031103 (2006), \texttt{gr-qc/0506126}

\bibitem{Frolov:2014jva}
V.P. Frolov, JHEP \textbf{05}, 049 (2014), \texttt{1402.5446}

\bibitem{Christensen:1977jc}
S.M. Christensen, S.A. Fulling, Phys. Rev. \textbf{D15}, 2088 (1977)

\bibitem{Frolov:1984}
V.P. Frolov, G.A. Vilkovisky, \emph{{Spherically Symmetric Collapse in Quantum
  Gravity}}, in \emph{{Quantum Gravity}} (Springer US, 1984), pp. 267--290

\bibitem{Vilkovisky:1985}
G.A. Vilkovisky, \emph{{The gospel according to DeWitt}}, in \emph{{Quantum
  theory of gravity}} (Adam Hilger Limited, Bristol UK, 1985), pp. 169--209

\bibitem{Poisson:1989zz}
E.~Poisson, W.~Israel, Phys. Rev. Lett. \textbf{63}, 1663 (1989)

\bibitem{Poisson:1990eh}
E.~Poisson, W.~Israel, Phys. Rev. \textbf{D41}, 1796 (1990)

\bibitem{Brown:2011tv}
E.G. Brown, R.B. Mann, L.~Modesto, Phys. Rev. \textbf{D84}, 104041 (2011),
  \texttt{1104.3126}

\bibitem{Bolashenko:1986mr}
P.A. Bolashenko, V.P. Frolov, in \emph{{Physical Effects in the Gravitational
  Fields of Black Holes , Proceedings of the Lebedev Physics Institute of the
  Academy of Sciences of the USSR, v.169}}, edited by M.A. Markov (Commack,
  N.Y., Nova Science, 1986), pp. 159--168

\end{thebibliography}
\end{document}